\documentclass[11pt,a4paper]{article}

\usepackage{epsfig}
\usepackage{graphicx}
\usepackage{epstopdf}
\usepackage[pdf]{pstricks}
\usepackage{amsmath}
\usepackage{amssymb}
\usepackage{verbatim}
\usepackage{bm}
\usepackage{dsfont}
\usepackage[footnotesize]{caption}
\usepackage{cite}
\usepackage{textcomp}
\usepackage{calc}
\usepackage{geometry}
\usepackage{bbm}
\usepackage{xcolor}
\usepackage[utf8]{inputenc}
\usepackage{enumitem}   
\usepackage{setspace}
\usepackage{upgreek}
\usepackage{amsmath}
\usepackage{slashed}
\usepackage{caption}
\usepackage{subcaption}

\newcommand{\iu}{i}
\newcommand{\ii}{i}

\usepackage{color}
\definecolor{cred}{RGB}{180,50,40} 
\definecolor{purple}{RGB}{180,90,180} 
\definecolor{darkgreen}{RGB}{0, 100, 0}
\definecolor{commentblue}{RGB}{0, 0, 255}

\usepackage{hyperref}
\hypersetup{colorlinks=true,urlcolor=blue,linkcolor=red,citecolor=green!60!black}

\begin{document}


\noindent September 2020
\hfill CERN-TH-2020-158, CP3-20-45, DESY 20-159

\vskip 0.4cm

\begin{center}
\bigskip
{\huge\bf
\begin{spacing}{1.1}
MeV-scale Seesaw and Leptogenesis
\end{spacing}
}

\vskip 0.4cm

\renewcommand*{\thefootnote}{\fnsymbol{footnote}}
{\large
Valerie Domcke\,$^{a\,\hspace{-0.25mm}}$, Marco Drewes  $^{b\,\hspace{-0.25mm}}$, Marco Hufnagel$^{c\,\hspace{-0.25mm}}$, and Michele Lucente $^{b,d\,\hspace{-0.25mm}}$
\\[3mm]
{\it{
$^{a}$
Theoretical Physics Department, CERN, 1 Esplanade des Particules, Geneva, Switzerland\\
Laboratory for Particle Physics and Cosmology,  EPFL, Lausanne, Switzerland}\\
$^{b}$
Centre for Cosmology, Particle Physics and Phenomenology, Universit\'e catholique de Louvain, Chemin du Cyclotron 2, B-1348 Louvain-la-Neuve, Belgium\\
$^{c}$
Deutsches Elektronen-Synchrotron (DESY), Hamburg, Germany\\
$^{d}$
Institute for Theoretical Particle Physics and Cosmology (TTK), RWTH Aachen University, D-52056 Aachen, Germany
}}
\end{center}
\vskip 0.4cm

\renewcommand*{\thefootnote}{\arabic{footnote}}
\newcommand{\axion}{a}
\setcounter{footnote}{0}


\begin{abstract}
We study the type-I seesaw model with three right-handed neutrinos and Majorana masses below the pion mass. In this mass range, the model parameter space is not only strongly constrained by the requirement to explain the light neutrino masses, but also by experimental searches and cosmological considerations. In the existing literature, three disjoint regions of potentially viable parameter space have been identified. In one of them, all heavy neutrinos decay shortly before big bang nucleosynthesis. In the other two regions, one of the heavy neutrinos either decays between BBN and the CMB decoupling or is quasi-stable. We show that previously unaccounted constraints from photodisintegration of nuclei practically rule out all relevant decays that happen between BBN and the CMB decoupling. Quite remarkably, if all heavy neutrinos decay before BBN, the baryon asymmetry of the universe can be quite generically explained by low-scale leptogenesis, i.e.\ without further tuning in addition to what is needed to avoid experimental and cosmological constraints. This motivates searches for heavy neutrinos in pion decay experiments.
\end{abstract}

\thispagestyle{empty}

\begin{small}
\tableofcontents
\end{small}

\newpage


\section{Introduction}\label{sec:intro}
The observation of neutrino-flavour oscillations is one of the few hints for new physics beyond the Standard Model that have been discovered to date. In fact, it is the only one found in the laboratory that has been established beyond doubt. One way of explaining these oscillations is by 
adding right-handed neutrinos $\nu_R$ to the Standard Model (SM) of particle physics, thus giving mass to the light neutrinos \cite{Minkowski:1977sc,GellMann:1980vs,Mohapatra:1979ia,Yanagida:1980xy,Schechter:1980gr,Schechter:1981cv}.
Such a \emph{type-I seesaw model} with right-handed neutrino masses below the electroweak scale is a  minimal and testable extension of the SM that can simultaneously explain the generation of the observed neutrino masses as well as the baryon asymmetry of the Universe. With the absolute mass scale of the SM neutrinos being bounded only from above, most studies in this context merely consider two right-handed neutrinos, thus leaving one SM neutrino massless. This drastically reduces the complexity of the problem, and in many cases serves as a good proxy for the relevant dynamics. However, based on both theoretical and experimental considerations, it is necessary to go beyond this simplification. 
Firstly, all other fermions in the SM come in three generations and overarching concepts such as gauging the difference of baryon and lepton number -- with a possible embedding in a grand unified theory -- mandate the introduction of three generations of right-handed neutrinos. 
Secondly, the combination of a high-dimensional parameter space together with neutrino oscillation data, constrains the theory to highly non-trivial sub-manifolds of the parameter space, where the naive intuition gained from the simplified model with only two neutrinos may fail. Finally, predicting particles at an energy scale within the reach of collider experiments, a selling point of this model is its falsifiability. To guide future experimental efforts, it is thus mandatory to map out the full range of potential observables, especially since it is well known that the inclusion of three right-handed neutrinos can significantly change the experimentally viable parameter space \cite{Chrzaszcz:2019inj}, cosmological constraints \cite{Hernandez:2014fha}, and the perspectives for leptogenesis \cite{Abada:2018oly}. In this work we study a comparably unexplored region of parameter space in which all heavy neutrinos have masses below the pion mass, kinematically limiting their decay products to SM neutrinos, electrons, positrons, and photons.

Suppressing SU(2) indices for brevity, the most general renormalisable Lagrangian including SM fields and the right-handed neutrinos $\nu_R$ reads
\begin{equation}
    \mathcal L
  = \mathcal L_{\rm SM} + \iu \overline{\nu_R}_i \slashed\partial \nu_{Ri}
  - \frac{1}{2} \left( \overline{\nu_R^c}_i(M_M)_{ij}\nu_{Rj}
    + \overline{\nu_R}_i(M_M^\dagger)_{ij}\nu_{R}^c \right)
  - F_{ai} \overline{\ell_{L a}} \varepsilon \phi^* \nu_{Ri}
  - F_{ai}^* \overline{\nu_R}_i \phi^T \varepsilon^\dagger \ell_{L a}
\, . \label{eq:Lagrangian}
\end{equation}
Here  $\ell_{L}$ and $\phi$ are the left-handed lepton and Higgs doublet of the SM, respectively, $F$ is the matrix of Yukawa couplings, and $\varepsilon$ denotes the totally antisymmetric SU(2) tensor.
The Majorana mass matrix $M_M$ introduces a new fundamental scale in nature, which is usually referred to as the \emph{seesaw scale}. More precisely, for $n$ flavours of $\nu_R$, Eq.~\eqref{eq:Lagrangian} contains $n$ new dimensionful parameters that can be identified with the eigenvalues of $M_M$, which roughly coincide with the physical masses $M_i$ of the heavy neutrino mass eigenstates $N_i$ (see Eq.~\eqref{eq:blocks_mass_matrix} below). The phenomenological and cosmological implications of the $\nu_R$'s existence strongly depend on the choice of the seesaw scale(s) (cf.\ e.g.\ \cite{Drewes:2013gca} for a review).

A particularly intriguing feature of this model is the fact that the same $\nu_R$ that give masses to the light neutrinos can also explain the observed matter-antimatter asymmetry in the early universe, which is believed to be the origin of all baryonic matter that is present today.\footnote{The evidence for a matter-antimatter asymmetry in the observable universe and its connection to the origin of matter are e.g.~discussed in ref.~\cite{Canetti:2012zc}.}
This is realized via the process of \emph{leptogenesis} \cite{Fukugita:1986hr}, which is feasible for a very wide range of possible $M_i$ (see Ref.~\cite{Bodeker:2020ghk} for a recent review). For $M_i$ above the electroweak scale, the asymmetry is typically generated during the freeze-out and decay of the heavy neutrinos \cite{Fukugita:1986hr} (``freeze-out scenario"), while for $M_i$ below the electroweak scale, it is instead generated during their production \cite{Akhmedov:1998qx,Asaka:2005pn,Hambye:2016sby} (``freeze-in scenario").\footnote{The statement that the freeze-out scenario works for $M_i$ above the electroweak scale and the freeze-in scenario works for $M_i$ below the electroweak scale should be thought of as a rule of thumb. In fact, both mechanisms overlap between roughly $\sim 5$ GeV
and the TeV scale \cite{Klaric:2020lov}.} 
It is well-known that leptogenesis is in principle feasible with $M_i$ in the range of a few MeV \cite{Canetti:2012kh}.
However, in this mass range, the model parameter space is strongly constrained by laboratory experiments, cosmology, and astrophysics.

Constraints on the properties of heavy neutrinos are conveniently expressed in terms of the mixing angles $\theta_{ai}$,
(cf.\ Eq.~\eqref{ThetaDef}
below).
In fact, for given $M_i$, the values of $\theta_{ai}$ determine the thermal $N_i$ production rate in the early universe, the $N_i$ lifetime, the $N_i$ contribution to the generation of light neutrino masses,
and the $N_i$ production cross-section in experiments.
For masses below $\sim 100$ MeV and values of $\theta_{ai}$ that are small enough to satisfy exclusion bounds from various laboratory experiments, the heavy neutrinos tend to have lifetimes larger than $0.1$~s.
This means that their presence in the primordial plasma and their decay may affect cosmological observables, such as the abundances of light elements that are produced during big bang nucleosynthesis (BBN), or the anisotropies in the cosmic microwave background (CMB). The resulting constraints on $M_i$ and $\theta_{ai}$ have e.g.~been summarised in \cite{Vincent:2014rja}.\footnote{
The authors in \cite{Vincent:2014rja} 
ruled out lifetimes longer than the CMB decoupling time by rescaling the CMB bounds on decaying Dark Matter particles found in \cite{Diamanti:2013bia}. These were obtained under the assumption that the particles have a lifetime that exceeds the age of the universe and can therefore strictly speaking not be applied in all of the parameter space considered here. However, it turns out that the parameter region where this rescaling is not applicable is ruled out by the results obtained in \cite{Poulin:2016anj}, so that we can safely apply the bounds presented in  \cite{Vincent:2014rja} here.
}
Usually, these limits can be avoided for sufficiently small values of $\theta_{ai}$. However, since the mixing angles also govern the size of the light neutrino masses, there exist additional lower bounds on different combinations of $\theta_{ai}$ from the requirement to explain the observed light neutrino oscillation parameters. 
These lower bounds depend on the number $n$ of right-handed neutrino flavours and the mass $m_{\rm lightest}$ of the lightest neutrino (cf.~\cite{Drewes:2019mhg} for a recent discussion).
In the minimal model with $n=2$ and $m_{\rm lightest}=0$, the seesaw mechanism necessarily enforces that all $N_i$ reach thermal equilibrium if their masses are below $\sim 100$ MeV \cite{Hernandez:2013lza}.
In combination with bounds from direct searches, this practically rules out the entire mass range below $\sim 100$ MeV ($\sim 350$ MeV) for normal (inverted) ordering of the light neutrino masses \cite{Drewes:2016jae}.\footnote{The authors of \cite{Drewes:2016jae} assumed a mass degeneracy among the $N_i$. However, since both, the lifetime bound from BBN and constraints from direct searches in good approximation apply to each $N_i$ individually, this can at most introduce a factor 2 in the upper bound on the mixing (if the two $N_i$ cannot be distinguished kinematically), which will not change these conclusions.\label{n2footnote}}

In the next-to-minimal model with $n=3$ considered here, one of the $N_i$ -- which we may call $N_1$ without loss of generality -- can have small enough mixings $\theta_{a1}$ to avoid equilibration in the early universe and hence is no longer constrained by the lower bound on $M_i$ if $m_{\rm lightest} \lesssim 10^{-3}$ eV \cite{Hernandez:2014fha}.\footnote{We do not consider the small window of $M_1$ in the eV range that was reported in \cite{Hernandez:2014fha} because the scenario of a eV seesaw \cite{deGouvea:2005er} is meanwhile even more disfavoured by cosmological date \cite{Aghanim:2018eyx}.}
This leaves three distinct regions of parameter space for models with $n=3$ and all $M_i$ below the pion mass. 
In \textbf{scenario I)} all three $N_i$ decay before BBN. A global fit of direct and indirect experimental constraints in this region has recently been performed in \cite{Chrzaszcz:2019inj}.
In \textbf{scenario II)} two of the $N_i$ decay before BBN. The third one never reaches thermal equilibrium and decays between BBN and the decoupling of the CMB.
In \textbf{scenario III)} two of the $N_i$ decay before BBN. The third one is quasi-stable and contributes to the Dark Matter  \cite{Dodelson:1993je,Shi:1998km}. This scenario corresponds to the well-known \emph{Neutrino Minimal Standard Model} ($\nu$MSM) \cite{Asaka:2005pn,Asaka:2005an}.

In the present work, we present two new results regarding these scenarios. Firstly, we demonstrate that scenario II) is ruled out when combining previously unaccounted constraints from photodisintegration after BBN with constraints from CMB anisotropies~\cite{Poulin:2016anj} and the ionisation of the intergalactic medium~\cite{Diamanti:2013bia}. 
Secondly, we find that the baryon asymmetry generated in scenario I) generically is of the right order of magnitude to explain the observed matter-antimatter asymmetry. This surprising result indicates that 
within the highly constrained region of parameter space where all experimental constraints are satisfied, no or little additional tuning is needed for successful leptogenesis. 
These results extend the previous parameter scan of scenario I) in \cite{Abada:2018oly} to smaller masses,  $M_i \gtrsim 50$ MeV. Finally, let us note that we do not consider baryogenesis in scenario III) and instead refer the reader to \cite{Boyarsky:2009ix,Canetti:2012kh} for a comprehensive overview and to \cite{Hernandez:2016kel,Drewes:2016jae,Antusch:2017pkq,Eijima:2018qke,Ghiglieri:2020ulj,Klaric:2020lov} for recent updates on the viable parameter space in this model. 
The remainder of this article is organised as follows. In Sec.~\ref{sec:constraints}, we summarise the existing laboratory and cosmological constraints, before introducing our new bound from photodisintegration after BBN in Sec.~\ref{sec:MarcoH}. We comment on the supernova bound in Sec.~\ref{sec:SNbound}, which could rule out the entire scenario I) but comes with some uncertainties. We summarise all constraints in Sec.~\ref{sec:viable}, demonstrating that the neutrino oscillation data can be accounted for in the remaining parameter space. Sec.~\ref{sec:BAU} is dedicated to the study of leptogenesis in scenario I), followed by a brief conclusion in Sec.~\ref{sec:discussion}.


\section{Laboratory, cosmological and astrophysical constraints}
\label{sec:constraints}

\subsection{Laboratory constraints}\label{sec:lab}

The strongest experimental constraints on the heavy neutrino properties come from the requirement to explain the light neutrino oscillation data.
If the eigenvalues of the Majorana mass matrix $M_M$ are at least a few eV in magnitude, there exist two distinct sets of mass eigenstates after electroweak symmetry breaking, which can be represented by the flavour vectors of Majorana spinors
\begin{eqnarray}\label{MassStates}
\upnu \simeq U_{\nu}^{\dagger} \left(\nu_{L} - \theta \nu_{R}^c\right) + {c.c.}
\quad , \quad
N \simeq U_N^\dagger \left( \nu_{R} +  \theta^T\nu_{L}^{c}\right) + {c.c.}\,.
\end{eqnarray}
Here $c.c.$ denotes the $c$-conjugation which e.g.~acts as $\nu_R^c=C\overline{\nu_R}^T$ with $C=\ii\gamma_2\gamma_0$,
$U_\nu$ is the standard light neutrino mixing matrix, and $U_N$ is its equivalent among the heavy neutrinos. 
The mixing between left- and right-handed neutrinos is quantified by the entries of the matrix 
\begin{equation}\label{ThetaDef}
\theta = v F M_M^{-1}\,,
\end{equation}
with the Higgs field expectation value $v$,
and the mass matrices for $\upnu$ and $N$ are given by
\begin{eqnarray}\label{eq:blocks_mass_matrix}
m_\nu = - \theta M_M \theta^T \ , \
M_N  = M_M + \frac{1}{2} (\theta^\dagger \theta M_M + M_M^T \theta^T \theta^{*})\,.\end{eqnarray}
The squares of the physical masses $m_i$ and $M_i$ of $\upnu_i$ and $N_i$, are given by the eigenvalues of the matrices $m_\nu^\dagger m_\nu$ and $M_N^\dagger M_N$. 
Here we work at tree level and expand all expressions to second order in the small mixing angles $\theta_{ai}$. 
The $\upnu_i$ can be identified with the well-known light neutrinos, while the $N_i$ are new heavy (almost) sterile neutrinos. Their masses $M_i$ coincide with the eigenvalues of $M_M$ up to $\mathcal{O}(\theta^2)$ corrections in Eq.~\eqref{eq:blocks_mass_matrix}. Within the pure seesaw model in Eq.~\eqref{eq:Lagrangian}, the $N_i$ interact with the SM only through their mixing with the doublet fields $\nu_L$ in Eq.~\eqref{MassStates}, which practically leads to a $\theta$-suppressed weak interaction.

The requirement to explain the observed light neutrino mass splittings $m_i^2-m_j^2$ as well as the mixing angles in the matrix $U_\nu$ imposes constraints on the matrix $m_\nu$, and therefore on $F$ and $M_M$. At low energies, this leads to restrictions on the relative size of the heavy neutrino mixing with individual SM flavours \cite{Shaposhnikov:2008pf,Ruchayskiy:2011aa,Asaka:2011pb,Hernandez:2016kel,Drewes:2016jae,Caputo:2017pit,Drewes:2018gkc,Chrzaszcz:2019inj}, i.e.\ on the quantities $U_{ai}^2/U_i^2$ with $U_i^2=\sum_a U_{ai}^2$ and
\begin{equation}\label{UaiDef}
U_{ai}^2=|\Theta_{ai}|^2 \quad {\rm with} \quad \Theta = \theta U_N^*\,.
\end{equation}
There also is a lower bound on the different $U_i^2$ from neutrino oscillation data \cite{Asaka:2011pb,Ruchayskiy:2011aa} which roughly reads $U_i^2 > m_{\rm lightest}/M_i$ (cf.~\cite{Drewes:2019mhg} for a recent discussion).

\begin{figure}
 \includegraphics[width= 0.46 \textwidth]{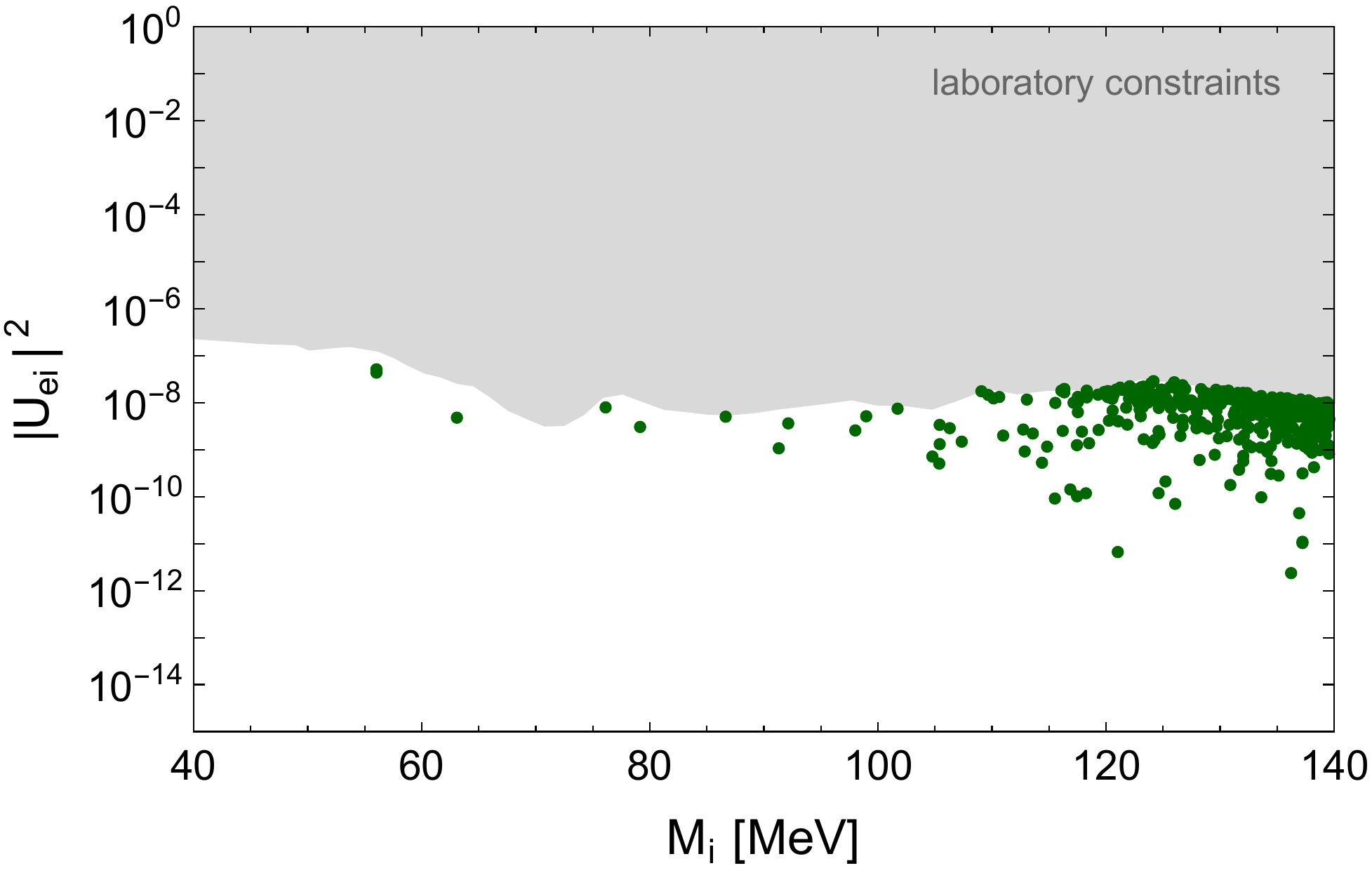} \hfill
  \includegraphics[width= 0.46 \textwidth]{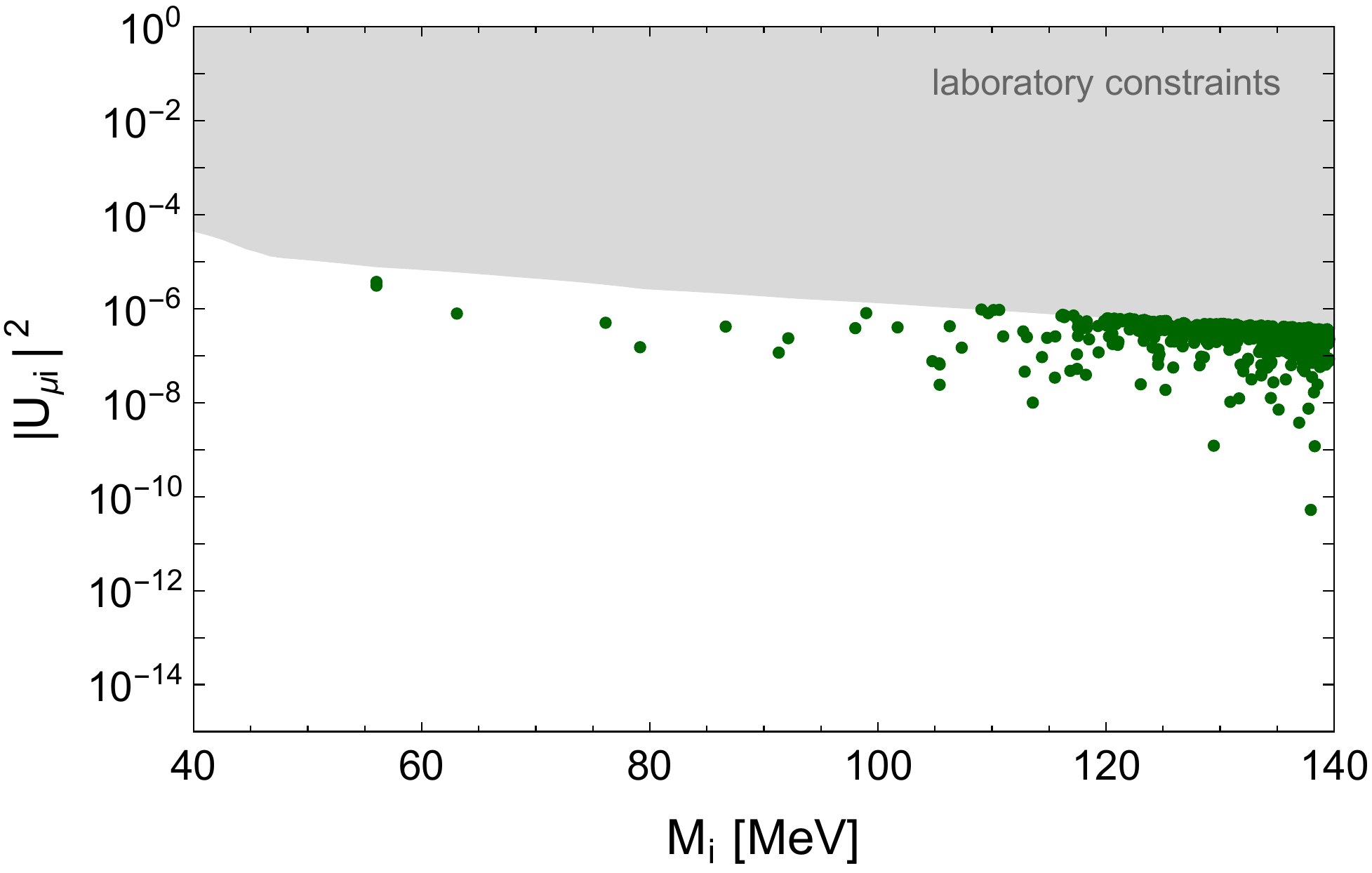}\vspace{0.4cm}
  \includegraphics[width= 0.46 \textwidth]{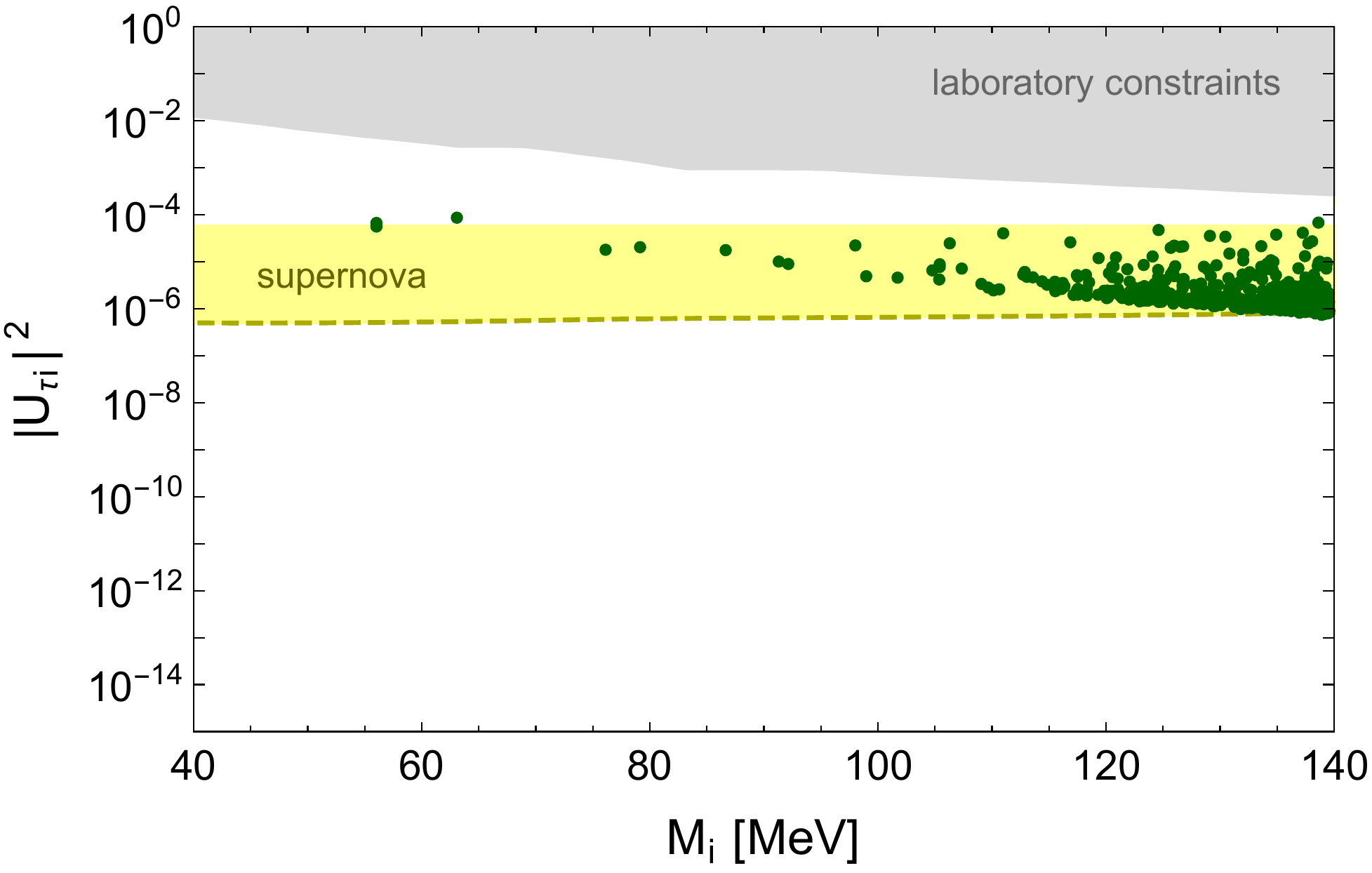} \hfill
  \includegraphics[width= 0.46 \textwidth]{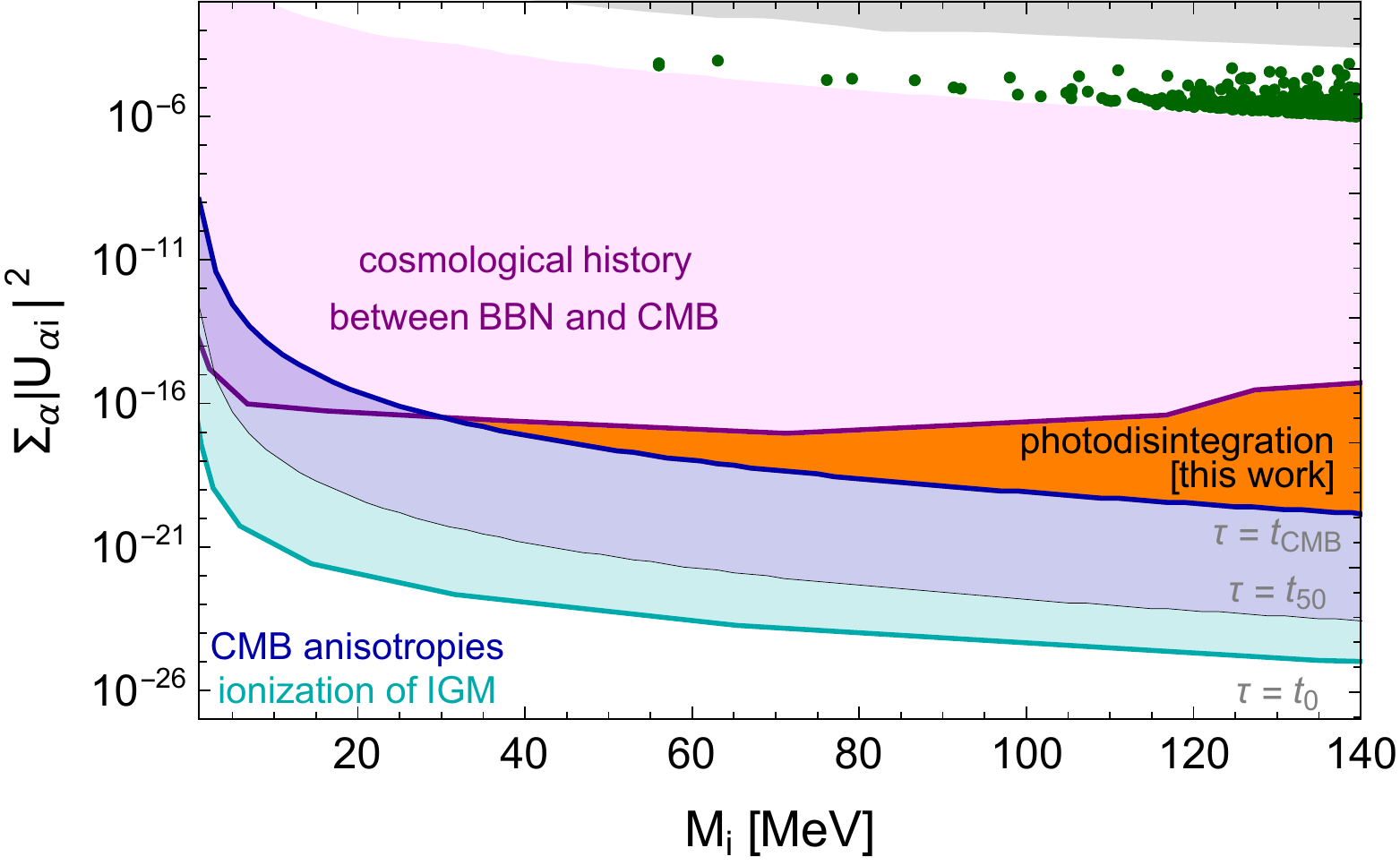}
\caption{Constraints on the sterile neutrino mixing with SM neutrinos. The shaded regions are excluded by laboratory constraints (grey, see Sec.~\ref{sec:lab}), cosmological constraints from Refs.~\cite{Vincent:2014rja} (pink), \cite{Poulin:2016anj} (blue) 
and \cite{Diamanti:2013bia} (cyan) (see Sec.~\ref{sec:cosmo}) and by the photodisintegration bound (orange, this work, Sec.~\ref{sec:MarcoH}). The supernova constraint from Ref.~\cite{Mastrototaro:2019vug} is indicated by the yellow shaded region (see Sec.~\ref{sec:SNbound}). The green dots show realisations of neutrino masses and mixings which reproduce the neutrino oscillation data (see Sec.~\ref{sec:viable}). Every parameter point is represented by a triplet (one point for each $N_i$).
}
\label{fig:mixing}
\end{figure}

The presence of weak interactions implies that a wide range of experiments is sensitive to the existence of the heavy neutrinos. An updated overview of the existing constraints that we are aware of can be found in \cite{Chrzaszcz:2019inj}. Broadly speaking, one can distinguish between direct and indirect searches. 
Direct searches are experiments in which the $N_i$ appear as real particles. 
If kinematically allowed, the $N_i$ production cross-section is roughly given by $\sigma_{N_i} \sim \sum_a U_{ai}^2\sigma_{\nu_a}$, with $\sigma_{\nu_a}$ being the production cross-section for a SM neutrino $\nu_a$. Hence, direct searches always impose upper bounds on the different $U_{ai}^2$. 
For sub-GeV masses this mainly includes beam dump experiments and peak searches.
Indirect searches include precision tests or searches for rare processes in the SM that are indirectly affected by the existence of the heavy neutrinos, e.g.~through the modification of the light neutrinos' interactions via the mixing $\theta$. In the mass range considered here, direct searches strongly dominate,\footnote{
For a more complete listing see the
\href{https://pdglive.lbl.gov/Particle.action?node=S077&init=0}{pdgLive page on HNLs} \cite{Zyla:2020zbs}.
} 
in particular from PIENU \cite{Aguilar-Arevalo:2019owf,Aguilar-Arevalo:2017vlf}, KEK \cite{Hayano:1982wu}, LBL \cite{Pang:1989ut}, SIN \cite{Abela:1981nf}, TRIUMF \cite{Britton:1992xv} and CHARM \cite{Orloff:2002de}
(cf.~also \cite{Abada:2017jjx,Bryman:2019ssi,Bryman:2019bjg}).
All of these constraints are summarized in the grey regions in Fig.~\ref{fig:mixing}.
The only indirect constraint that is relevant in this region comes from neutrinoless double $\beta$-decay ($0\nu\beta\beta$). However, the rate of the $0\nu\beta\beta$-decay can be suppressed even for mixing angles that are orders of magnitude larger than the ones considered here if one requires that the Lagrangian in Eq.~\eqref{eq:Lagrangian} approximately conserves a generalisation of the SM lepton number $L$ (more precisely, the difference between baryon number $B$ and $L$) under which the heavy neutrinos are charged \cite{Shaposhnikov:2006nn,Kersten:2007vk},\footnote{In Ref.~\cite{Kersten:2007vk} it was pointed out that imposing a generalised $B-L$ symmetry can lead to a parametric suppression of all lepton number violating observables.
This suppression indeed happens for the $0\nu\beta\beta$-decay and for the (Majorana) masses $m_i$ of the light neutrinos (where it is necessarily needed to allow for mixings $U_i^2\gg \sqrt{\sum_j m_j^2}/M_i$ without tuning).
However, in the mass range considered here, this symmetry does not suppress lepton number violating signatures in collider based experiments \cite{Drewes:2019byd}. 
}
and the current bound on the  $0\nu\beta\beta$-lifetime rules out only a small fraction of the leptogenesis parameter space \cite{Drewes:2016lqo,Hernandez:2016kel,Abada:2018oly}.

\subsection{Summary of previously known cosmological constraints}\label{sec:cosmo}

In the following we assume that the $N_i$ have masses of at least a few MeV. For lower $M_i$ the analysis would have to take into account the fact that the $N_i$ are still partially relativistic when the SM neutrinos decouple from the primordial plasma, which would require a more careful analysis of BBN. 
This will generally lead to even stronger bounds than the ones considered here because the $N_i$ would e.g.~contribute to $N_{\rm eff}$. 
Sterile neutrinos in the ${\cal O}(10 - 100)$~MeV mass range can alter our cosmological history and are hence strongly constrained by observations related to BBN and the CMB. Here we distinguish three cases, depending on the lifetime $\tau_i$ of the sterile neutrino $N_i$ \cite{Atre:2009rg},
\begin{equation}\label{LifeTime}
 \tau_{i}^{-1} \approx 7.8 \, \text{s}^{-1} \left(\frac{M_i}{10~\text{MeV}}\right)^5 \left[1.4 \ U_{ei}^2 +  U_{\mu i}^2 + U_{\tau i}^2 \right]\,.
\end{equation}

\paragraph{(i) Short-lived $N_i$.} If $N_i$ decays significantly before BBN, its decay products are fully thermalised and merely lead to a shift in the overall temperature of the thermal bath, which only shifts the onset of BBN. Hence, the highly constrained process of nucleosynthesis as well as the post-BBN cosmic history remain largely unaltered. 
This condition results in an upper bound on the lifetime of $N_i$ of ${\cal O}(0.2 - 1)$~s for 30~MeV$\leq M_i \leq $140~MeV~\cite{Dolgov:2000pj,Dolgov:2000jw,Ruchayskiy:2012si,Sabti:2020yrt,Boyarsky:2020dzc}.
Such short lifetimes require a sizeable mixing with SM neutrinos (above the region labeled `\emph{cosmological history between BBN and CMB}' in Fig.~\ref{fig:mixing}), which leads to a non-trivial interplay with the laboratory constraints discussed above (gray region in Fig.~\ref{fig:mixing}).\footnote{Note that these cosmologically ``short-lived" $N_i$ are still classified as ``long-lived particles" from the viewpoint of accelerator-based experiments. Their decay length e.g.~exceeds the size of the LHC main detectors.}

\paragraph{(ii) Long-lived $N_i$.} Heavy neutrino decays during BBN would directly alter the formation of light elements. 
If $N_i$ decays after BBN (but before CMB decoupling) it can impact the post-BBN cosmological history. In extreme cases, the non-relativistic sterile neutrinos can even come to dominate the energy budget of the Universe. Moreover, their decay leads to an entropy injection into the SM thermal bath. This leads to an upper bound on the mixing $\sum_\alpha |U_{\alpha i}|^2$ of $N_i$ with the SM neutrinos $\nu_\alpha$~\cite{Vincent:2014rja} (region labeled `\emph{cosmological history between BBN and CMB}' in Fig.~\ref{fig:mixing}).\footnote{Note that the setup of Ref.~\cite{Vincent:2014rja} contains only one sterile neutrino which couples exclusively to $\nu_e$. Taking into account the actual flavour structure in the couplings, we impose the bound derived in Ref.~\cite{Vincent:2014rja} on the mixing summed over all SM flavours.}
A further constraint arises from the effective number of relativistic degrees-of-freedom $N_\text{eff}$ at the time of BBN. As was demonstrated in Ref.~\cite{Hernandez:2014fha}, at least two out of the three sterile neutrinos temporarily reach thermal equilibrium - and consequently a sizeable abundance - in the early Universe.  This leads to a significant contribution to $N_\text{eff}$ during BBN if the sterile neutrino is relativistic at decoupling. 
We find the resulting upper bound on the mixing between active and sterile neutrinos to be weaker than the constraint derived in~\cite{Vincent:2014rja} in the parameter space of interest.

\paragraph{(iii) Quasi-stable $N_i$.}
$N_i $ lifetimes beyond the time of CMB decoupling ($\sim 10^{12}$ s) are highly constrained by CMB observations \cite{Slatyer:2016qyl,Poulin:2016anj} (region labeled `\textit{CMB constraints}' in Fig.~\ref{fig:mixing}), the impact of their decays on the intergalactic medium (IGM) \cite{Diamanti:2013bia,Acharya:2019uba} (region labeled `\textit{IGM constraints}' in Fig.~\ref{fig:mixing}), and the produced X-rays \cite{Lattanzi:2013uza}. 
The constraints can be avoided for sufficiently small $\theta_{ai}$, which suppresses both the thermally produced abundance and the $N_i$ decay rate.
In the mass range considered here, such a long lifetime requires mixing angles that are so tiny that the amount of thermally produced $N_i$ is negligible for all practical purposes (white region at the bottom of Fig.~\ref{fig:mixing}).\footnote{ 
For masses in the keV range, Eq.~\eqref{LifeTime}
permits mixing angles that are large enough that thermally produced $N_i$ can make up a considerable fraction of the DM and the bounds summarised in  \cite{Adhikari:2016bei,Boyarsky:2018tvu} should be applied.}

\medskip
This leads to the three distinct regions I)-III) of the parameter space which survive both the laboratory and cosmological constraints. 
In scenario I) all three sterile neutrinos belong to population~(i). 
They have relatively large mixing angles, and thermalise and decay before BBN. This region is found by applying the bound on the lifetime of sterile neutrinos from Refs.~\cite{Ruchayskiy:2012si,Gelmini:2020ekg,Sabti:2020yrt,Boyarsky:2020dzc}.
In scenario II), two of the heavy neutrinos $N_2$ and $N_3$ belong to population (i).
The third heavy neutrino $N_1$ features significantly smaller mixings $\theta_{a1}$ with the SM states and belongs to population~(ii).
$N_1$  avoids thermalisation~\cite{Hernandez:2014fha}
and obeys the bounds derived in Ref.~\cite{Vincent:2014rja}. 
More precisely, we use the bound depicted in Fig.~2 of Ref.~\cite{Vincent:2014rja},
which leaves open a window for $M_1 \gtrsim 50$ MeV and $|\theta_{ai}|^2 < 10^{-14}$.
However, as we will see in the following Sec.~\ref{sec:MarcoH}, this window is closed if the effect of $N_i$ decays on photodisintegration of nuclei is taken into account.
Scenario III) is similar to scenario II), but $N_1$ has even smaller mixings and is part of population (iii).

\subsection{Additional constraints from photodisintegration of nuclei}\label{sec:MarcoH}
Further, the $N_i$ decay can also disintegrate nuclei in the primordial plasma after BBN. The resulting bound strongly depends on the hadronic branching ratio of the decay, which vanishes for the mass range considered here, and does not affect any of the points in our sample \cite{Jedamzik:2006xz}. Hence, we only have to take into account electromagnetic decay channels, which we incorporate via the procedure described in \cite{Hufnagel:2018bjp, Depta:2020zbh} by running the public code \texttt{ACROPOLIS} \cite{Depta:2020mhj}.\footnote{Decays into SM neutrinos do not lead to photodisintegration and therefore can be neglected.} On that note, we first determine the non-thermal photon/electron-spectra by solving the full cascade equation \cite{Kawasaki:1994sc} with the appropriate source terms $\propto n_{i}/\tau_{i}$. The resulting spectra are then used to determine the late-time modifications of the nuclear abundances via photodisintegration by solving the appropriate non-thermal Boltzmann equation. Finally, we compare the resulting abundances with the most recent set of observations \cite{pdg,Geiss2003}. Specifically, we use
\begin{align}
\mathcal{Y}_\mathrm{p} &= (2.45 \pm 0.03) \times 10^{-1} \label{eq:Yp_abundance} \,,\\
\text{D}/{}^1\text{H} &= (2.547 \pm 0.025) \times 10^{-5} \label{eq:D_abundance} \,,\\
{}^3\text{He}/\text{D} &= (8.3 \pm 1.5) \times 10^{-1} \label{eq:3HeH_abundance} \,.
\end{align}
The resulting constraints are shown in the lower right panel of Fig.~\ref{fig:mixing} (orange) and we find that these additional limits are particularly important for closing the region of parameter space between the solid blue and purple line, i.e.\ the region that is otherwise neither excluded by CMB observations nor by a modified cosmological history between BBN and CMB.

\subsection{Supernovae bound}\label{sec:SNbound}
The detection of SN 1987A neutrinos arriving over an interval of about 10~s, in agreement with the predictions of a core-collapse supernova with the standard cooling scenario, imposes constraints on the existence of light BSM particles which would constitute an additional channel of energy-loss, shortening the duration of the neutrino burst~\cite{Raffelt:1990yz}. This has in particular been used to constrain axions~\cite{Raffelt:1990yz}, dark photons~\cite{Chang:2016ntp}, and sterile neutrinos of different mass ranges~\cite{Raffelt:2011nc,Dolgov:2000pj,Dolgov:2000jw,Mastrototaro:2019vug,Arguelles:2016uwb,Syvolap:2019dat,Suliga:2020vpz}. The constraints are particularly relevant for the mixing with $\nu_\tau$, since the laboratory constraints are weakest in this case. We indicate the constraints found in~\cite{Mastrototaro:2019vug} by the yellow shaded area in the bottom left panel of Fig.~\ref{fig:mixing}. 

However, as it has been recently pointed out in Ref.~\cite{Bar:2019ifz}, these bounds rely on the standard core-collapse supernova model. If instead the supernova is modelled by a collapse-induced thermo-nuclear explosion~\cite{Blum:2016afe}, the observed neutrino signal could stem from the accretion disk and would be insensitive to the cooling rates. With this in mind, we do not apply the supernova bounds of Ref.~\cite{Mastrototaro:2019vug} in our main analysis, but emphasize that this region of parameter space can be fully probed in the near future - both by laboratory and astrophysical observations.

\subsection{Viable parameter space}\label{sec:viable}

The photodisintegration bound introduced in Sec.~\ref{sec:MarcoH} excludes all points of type (ii) in the mass range considered here and therefore rules out scenario II). As already stated in the introduction, the phenomenology of scenario III) corresponds to that of the much-studied $\nu$MSM and shall not be further investigated here. This leaves us with scenario I).
A priori it is not clear whether there are any viable parameter choices for which all bounds can be fulfilled \emph{simultaneously}. 
This is non-trivial because neutrino oscillation data restricts the flavour mixing pattern, i.e.\
the range of allowed values for $U_{ai}^2/U_i^2$, meaning that it may not be possible to fit all three $N_i$ into the allowed (white) parameter regions in  Fig.~\ref{fig:mixing}. 
It is well-known that this considerably constrains the range of allowed masses $M_i$ below the kaon mass in scenario III),
as the constraints on $N_2$ and $N_3$ in this scenario are practically identical to those in the model with only two heavy neutrinos because $N_1$ cannot make a measurable contribution to the seesaw mechanism.\footnote{
This can be seen by inserting the largest mixing angels for population (iii) in figure \ref{fig:mixing} into the seesaw formula \eqref{eq:blocks_mass_matrix}. An estimate $m_{\rm lightest} \sim U_1^2 M_1$ yields values $m_{\rm lightest} < 10^{-12} {\rm eV}$ for  $M_i$ in the MeV range, in which case the constraints on $N_2$ and $N_3$ are practically identical to those in the minimal model with two heavy neutrinos, cf.~figure 11 in \cite{Chrzaszcz:2019inj}. Bearing in mind the caveat already pointed out in footnote \ref{n2footnote} the results found in section 2 of \cite{Drewes:2016jae} can therefore be applied to scenario III).
In the $\nu$MSM larger values of $U_1^2$ are allowed because $M_1$ is in the keV range, but a similar conclusion can be drawn  \cite{Boyarsky:2006jm}.
} 
 When combined with direct search data and bounds from BBN, this practically rules out most heavy neutrino masses below $\sim 350$ MeV, with a few small windows between $100$ MeV and $300$ MeV left open if the light neutrino mass ordering is normal \cite{Drewes:2016jae}.\footnote{
The authors of \cite{Drewes:2016jae} used the BBN bounds from \cite{Ruchayskiy:2012si} in their global analysis. The updated bounds on $\tau_i$ from \cite{Boyarsky:2020dzc} are stronger for $M_i$ between the pion and kaon masses, and including them in a global analysis is likely to close some of these windows.}

For scenario I) this question has been studied in \cite{Chrzaszcz:2019inj}, where it was found that the combination of all experimental and cosmological bounds indeed leaves a sizeable region of viable parameter space with $M_i$ below the pion mass. The reason is that neutrino oscillation data in this case allows for larger values of $U_{\tau i}^2/U_i^2$ than in scenario III); this permits the heavy neutrinos to decay through their mixing with the third SM generation before BBN while respecting the direct search bounds on $U_e^2$ and $U_\mu^2$, which are much stronger than those on $U_\tau^2$.
Hence, HNLs with masses well below the kaon mass should be ``tau-philic".

However, the analysis in  \cite{Chrzaszcz:2019inj} did not include the supernova bound discussed in Sec.~\ref{sec:SNbound}. 
A complete scan of the allowed parameter region is numerically extremely expensive because of the high dimensionality of the parameter space (18 free parameters) and the complicated shape that the sub-manifolds defined by the various experimental constraints in the mass region considered here form in this space. 
Instead, we perform a limited scan with 
randomised parameter choices.
We use the radiatively corrected \cite{Lopez-Pavon:2015cga} Casas-Ibarra parameterisation \cite{Casas:2001sr}. For the mass splittings and the complex angles in the Casas-Ibarra parameterisation, we alternate between drawing our parameters from a linear versus a logarithmic distribution, as in Ref.~\cite{Abada:2018oly}.
We apply all experimental and cosmological constraints summarised above.
For the experimental bounds, we use the simple strategy adapted in Refs.~\cite{Drewes:2016jae,Drewes:2015iva} and interpret the exclusion regions published by the experimental collaborations as hard cuts (rather than using full likelihood functions as in Ref.~\cite{Chrzaszcz:2019inj}), which is sufficient for the purpose of this work.
For the lifetime constraints from BBN we use the results from \cite{Ruchayskiy:2012si}.\footnote{The more recent bounds from \cite{Sabti:2020yrt} agree with those. 
In \cite{Boyarsky:2020dzc} it was pointed out that the bound on $\tau_i$ becomes considerably stronger when one includes the effect of mesons that are produced in heavy neutrino decays on the primordial light element abundances. However, this does not apply to the mass range considered here because the heavy neutrinos can only decay into purely leptonic final states.}
We show a representative set of viable parameter points (indicated by green dots) that are consistent with all experimental and cosmological constraints in Fig.~\ref{fig:mixing}. These all correspond to the normal ordering of the SM neutrinos and require $U_{\tau i}^2 > U_{e i}^2, U_{\mu i}^2$. 

Taking into account that each parameter point is represented by a triplet of points in Fig.~\ref{fig:mixing} (one for each sterile flavour), applying the supernova bound from Ref.~\cite{Mastrototaro:2019vug} would exclude all points shown.  
However, as pointed out in Sec.~\ref{sec:SNbound}, this bound strongly relies on the underlying model for the supernova explosion.



\section{Baryogenesis}
\label{sec:BAU}

We now proceed to compute the baryon asymmetry for all viable parameter points found in our scan, using the set of quantum kinetic equations given in Sec.~2 of Ref.~\cite{Abada:2018oly} to describe the evolution of the heavy neutrino abundances and lepton asymmetries in the early universe.\footnote{The momentum dependent sets of kinetic equations derived in Refs.~\cite{Ghiglieri:2019kbw,Bodeker:2019rvr} are more accurate than the momentum averaged equations used in Ref.~\cite{Abada:2018oly}, but require a much larger numerical effort. Since the results are typically comparable \cite{Ghiglieri:2017csp} we opted for the simpler approach in the present work.} 

We assume that the radiation dominated epoch of the cosmic history started with a matter-antimatter symmetric primordial plasma in which all SM particles were in thermal equilibrium at a temperature $T_R$ that was much hotter than the temperature $T_{\rm sph}\simeq 131$ GeV \cite{DOnofrio:2014rug}, above which electroweak sphalerons efficiently convert $L$ and $B$ into each other \cite{Kuzmin:1985mm}. In inflationary cosmology, this is expected because pre-inflationary asymmetries would be diluted very efficiently by the cosmic expansion. We moreover take  the initial abundance of the heavy neutrinos to be negligible. The Lagrangian in Eq.~\eqref{eq:Lagrangian} then contains all the necessary ingredients to generate the baryon asymmetry of our Universe:
 The heavy neutrinos are generated from thermal interactions in the plasma through their 
 Yukawa couplings. 
 In this out-of-equilibrium situation, the interplay of coherent neutrino oscillations and decoherent scatterings mediated by the $CP$-violating Yukawa couplings can generate a lepton asymmetry that is partially converted into a baryon asymmetry by the sphalerons.
 For the $M_i$ under consideration here, and in view of the experimental constraints on the $U_i$, this process happens very slowly. If at least one heavy neutrino has not reached thermal equilibrium at $T\sim T_{\rm sph}$, then the baryon asymmetry is preserved (``frozen in") at lower temperatures. 
This \emph{freeze-in leptogenesis} mechanism, also known as Akhmedov-Rubakov-Smirnov (ARS) leptogenesis~\cite{Akhmedov:1998qx}, has been studied by many authors, a review is e.g.~given in Ref.~\cite{Drewes:2017zyw}.
Our goal is to study the question of whether the observed baryon asymmetry of the Universe can be explained in scenario I)
while respecting the constraints discussed in Sec.~\ref{sec:constraints} if all $M_i$ are smaller than the pion mass.

\begin{figure}
\centering
 \includegraphics[width = 0.7 \textwidth]{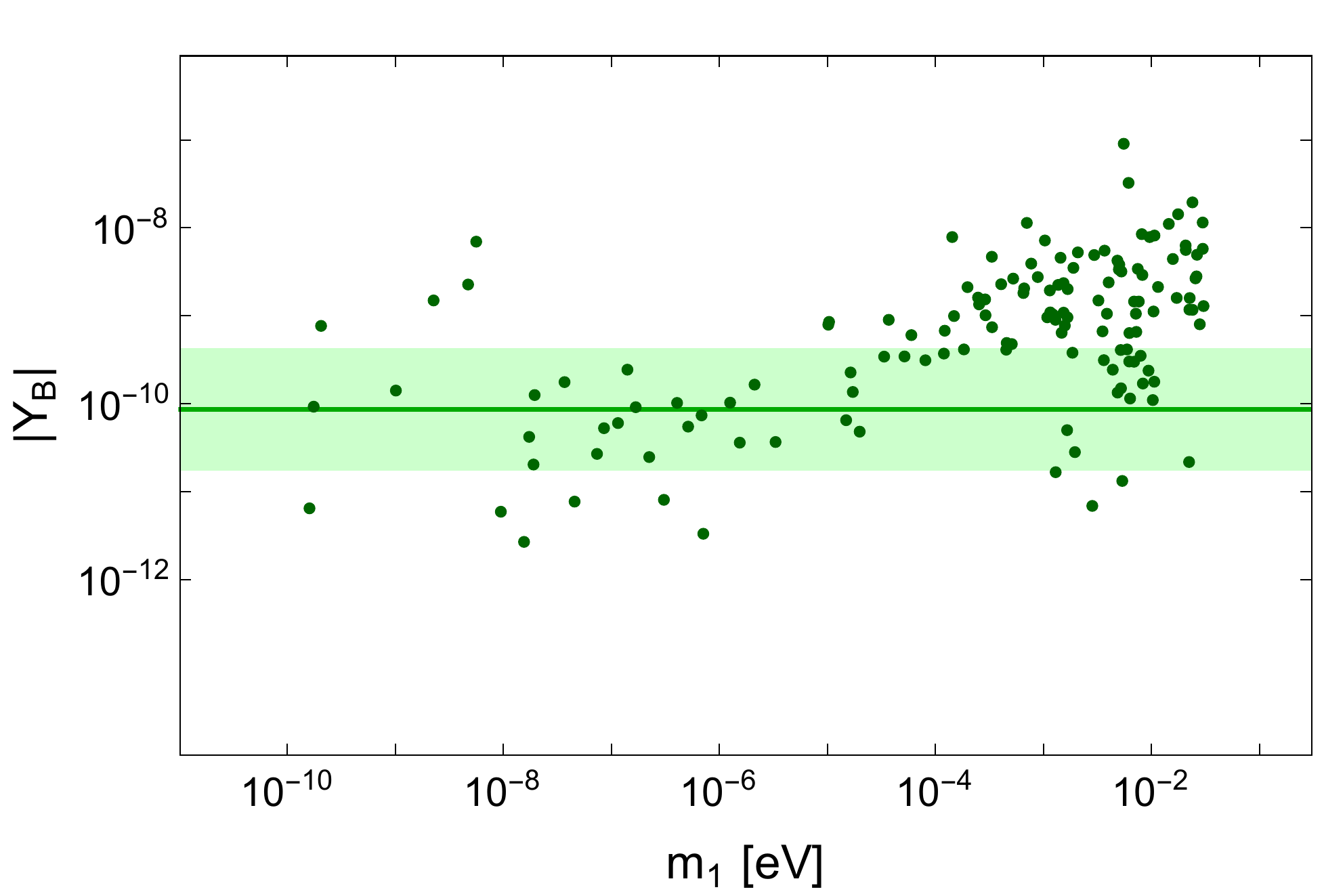}
 \caption{Predicted baryon asymmetry as a function of the lightest SM neutrino mass in scenario I). The green line indicates the observed value, the shaded region indicates an order of magnitude variance.
 }
 \label{fig:BAU}
\end{figure}

Our results are shown in Fig.~\ref{fig:BAU}. 
Remarkably, if we consider the population where all three sterile neutrinos decay before BBN, the predicted baryon asymmetry is generically in the correct ball-park to explain the observed value or larger. This is far from trivial since it is well known that marginal changes in the model parameters can lead to drastic changes in the resulting baryon asymmetry, due to the fine balance between generation and wash-out of the asymmetries.

\section{Discussion and conclusion}\label{sec:discussion}

We study the type-I seesaw model with three heavy neutrinos $N_i$ with masses between a few MeV and the pion mass. 
This part of the parameter space is relatively little studied because it is ruled out by the combination of direct searches for heavy neutrinos, cosmological constraints and light neutrino oscillation data if one considers the minimal seesaw with only two heavy neutrinos or the $\nu$MSM, where neutrino oscillation data prohibits large hierarchies between the mixings $U_{ai}^2$ with individual SM generations $a=e,\mu,\tau$. 
However, in a general framework with three right-handed neutrinos, neutrino oscillation data permits a comparably large mixings with the third SM generation that allows the $N_i$ to decay before BBN while respecting the tighter direct search constraints on the mixings with the first two generations. 
There are three scenarios with all heavy neutrino masses $M_i$ below the pion mass that are potentially allowed by previously published constraints: I) all three $N_i$ decay before BBN, II) two $N_i$ decay before BBN and the third one decays between BBN and the CMB decoupling, and III) two $N_i$ decay before BBN and the third one has a lifetime that greatly exceeds the age of the universe. 

Scenario III) can be ruled out when all $M_i$ are in the MeV range by noticing that the allowed mixing angles for the quasi-stable $N_1$ are so tiny that this particle effectively decouples, and the constraints on $N_2$ and $N_3$ are practically identically to those in the minimal model with two heavy neutrinos, where this mass range is known to be ruled out. 
We further find that scenario II) is ruled out by the effect that photons produced in a cascade from the long-lived $N_i$ decay would have on the disintegration of light elements in the IGM.
In scenario I) a representative randomised parameter scan shows that there are viable parameter values for which the $N_i$ can avoid all constraints from laboratory experiments and cosmology for $M_i>50$ MeV. 
All these points can potentially be ruled out by the observed neutrino flux from the supernova 1987a, but this conclusion depends on the modelling of the supernova explosion. 
Quite surprisingly most of the viable points give a final baryon asymmetry in the correct ball-park to explain the observed value. Given the well-known strong sensitivity of the relevant Boltzmann equations to small changes in the parameters, this is a highly non-trivial result.

Our results show that heavy neutrinos masses down to a few tens of MeV are allowed by experimental and cosmological constraints if all of them decay well before the onset of BBN in the early universe, and if they primarily mix with the third SM generation.
$N_i$ in this mass range can be found in the decay of light mesons, in particular pions, which are produced in large numbers in accelerator based searches. While the LHC main detectors have no sensitivity in this mass range due to both, the low transverse momentum of the events and the long lifetime of the $N_i$, dedicated detectors in the forward directions, such as FASER \cite{Feng:2017uoz}, instrumentation in the beam pipe \cite{SnowmassInstrumendetBeamPipeLoI} or the Forward Physics Facility \cite{SnowmassFPFLoI}, could potentially search for such $N_i$. Moreover, the mass range below the pion mass can be probed by fixed target experiments. For instance, the neutrino beams of DUNE \cite{Acciarri:2015uup} and T2K \cite{Abe:2011ks} contain a fraction of $N_i$, \footnote{
The fraction of pions decaying into $N_i$ can be estimated as 
${\rm B}(\pi^+\to e^+ N_i)\simeq U_{ei}^2\times{\rm B}(\pi^+\to e^+\nu_e)\simeq  10^{-4} U_{ei}^2$ if the heavy neutrino mass is neglected. The decay into $\mu^+ N_i$ does not suffer from the suppression factor $10^{-4}$, but is kinematically forbidden in the mass range where we found viable points. Kaons that contaminate the pion beam can be another source of $N_i$.
} 
the decay of which (e.g.~$N_i \to e^+ e^- \nu_a$) could be observed in their near detectors 
\cite{Coloma:2020lgy,Ballett:2019bgd}
or with additional detectors \cite{Arguelles:2019xgp}. The channel with $a=\tau$ can be used to probe the coupling $U_{\tau i}^2$ to the third generation, which tends to be the largest to ensure decay BBN while respecting the stronger experimental bounds on $U_{e i}^2$ and $U_{\mu i}^2$.
It would also be worth to investigate searches at the existing SBL \cite{Antonello:2015lea} and NuMi \cite{Adamson:2015dkw} beam lines at FNAL.

This work provides a proof-of-existence for a viable leptogenesis scenario in this mass range, but this is by no means an exhaustive study. Due to the high-dimensional parameter space, this requires more sophisticated numerical techniques, but we hope that the results presented here will trigger further work in this direction. This will be crucial in guiding experimental effort in fully testing freeze-in leptogenesis as the mechanism to generate the baryon asymmetry of our Universe.


\paragraph{Acknowledgements}
We thank Albert De Roeck, Anastasiia Filimonova,  Gaia Lanfranchi,
Laura Lopez-Honorez,
Nashwan Sabti,
Filippo Sala, Misha Shaposhnikov, Anna Sfyrla and Yun-Tse Tsai 
for very helpful discussions.
This work was partially funded by the ERC Starting Grant
‘NewAve’ (638528) as well as by the Deutsche Forschungsgemeinschaft (DFG) under Germany's Excellence Strategy - EXC 2121 Quantum Universe - 390833306.
M.L. acknowledges partial support from the Alexander von Humboldt Foundation.


\bibliographystyle{utphys}
\bibliography{refs}{}

\end{document}